\documentclass[12pt]{iopart}
\makeatletter
\@namedef{ver@amsmath.sty}{}
\makeatother
\usepackage{amstext}
\usepackage[latin1]{inputenc}
\usepackage{amsmath}
\usepackage{amsfonts}
\usepackage{amssymb}
\usepackage{graphicx}
\usepackage{xcolor}

\begin{document}

\title{Prospects for strongly coupled atom-photon quantum nodes}


\author{N Cooper$^{1,}$*, C Briddon$^1$, E Da Ros$^1$, V Naniyil$^1$, M T Greenaway$^2$ \& L Hackermueller$^1$}

\address{$^1$School of Physics and Astronomy, University of Nottingham, Nottingham NG7 2RD \\$^2$Department of Physics, Loughborough University, Loughborough LE11 3TU \\ *ppznc@nottingham.ac.uk}
\vspace{10pt}

\section{ABSTRACT}
We discuss the trapping of cold atoms within microscopic voids drilled perpendicularly through the axis of an optical waveguide. The dimensions of the voids considered are between 1 and 40 optical wavelengths. By simulating light transmission across the voids, we find that appropriate shaping of the voids can substantially reduce the associated loss of optical power. Our results demonstrate that the formation of an optical cavity around such a void could produce strong coupling between the atoms and the guided light. By bringing multiple atoms into a single void and exploiting collective enhancement, cooperativities $\sim 400$ or more should be achievable. The simulations are carried out using a finite difference time domain method. Methods for the production of such a void and the trapping of cold atoms within it are also discussed.
%
%

\section{Introduction}

The introduction of cold atoms into microscopic holes in optical waveguides allows the integration of atomic components into otherwise purely photonic devices, with potential applications in sensing and quantum information processing \cite{hinds,Reiserer2015}. While alternative techniques are available for coupling guided light to cold atoms --- for example the use of tapered nanofibres \cite{nanofib1,nanofib2,nanofib3} or hollow core fibres \cite{hcore1,hcore2,hcore3} --- microscopic holes offer a unique set of advantages that make them ideally suited for certain applications. Firstly, the technique of introducing cold atoms via a microscopic hole is just as applicable in a 2D waveguide chip as in a fibre, allowing the direct combination of cold atoms with photonic circuit devices. Secondly, while the overall optical depth of an atom cloud contained in a microscopic hole is likely to be less than that obtained using a nanofibre or hollow core fibre, the optical depth per unit length should be able to match that achievable in free space, which is substantially greater than that typical of hollow core fibre or nanofibre experiments. This may have important implications for spatial resolution in sensing applications. Finally, the spatial separation between the atoms and the solid material of the waveguide can be much larger in a microscopic hole than is typical of a hollow core fibre or possible using a nanofibre. This will be important in precision sensing and spectroscopy experiments, where atom-surface interactions might otherwise adversely affect the results, as well as for any experiments involving Rydberg atoms, which are currently one of the most promising candidates for the implementation of multi-qubit gates \cite{rberg1,rberg2,rberg3}.

\begin{figure}
 \begin{center}
 \includegraphics[width = \textwidth]{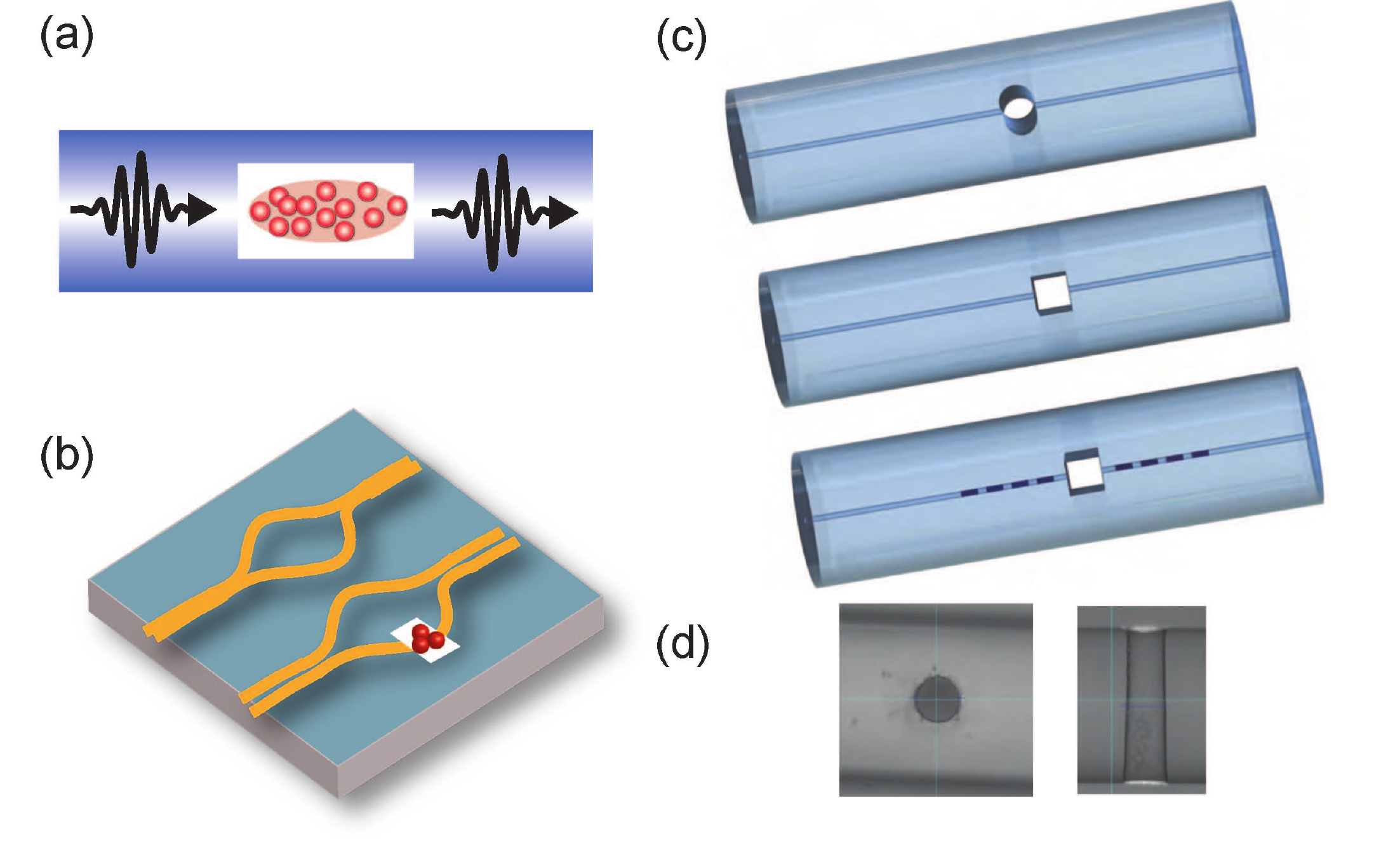} 
 \hspace{0.5cm}
 \caption{a) Illustration showing atoms confined within a microscopic void in an optical fibre, interacting with the guided light. b) The same method can be used to introduce cold atoms into a waveguide on a chip, thus allowing them to be integrated into more complex photonic circuits. c) Examples of possible geometries; round hole, square hole, square hole combined with fibre Bragg gratings. d) Laser drilled holes in a Thorlabs 780HP optical fibre, images produced by Workshop of Photonics.}
 \label{holepics}
 \end{center}
\end{figure}

In this article we simulate transmission of light across microscopic holes in optical waveguides. We focus on holes drilled perpendicular to the core of the waveguide, with diameters in the range of 1 to 40 optical wavelengths. Holes of this kind can be fabricated by pulsed laser drilling \cite{ldrill1,ldrill2} --- see for example those shown in figure \ref{holepics}(d), which were fabricated by Workshop of Photonics \cite{wop} in 2015. We find that appropriate shaping of the holes can significantly improve the overlap of the transmitted light with the guided mode, thus reducing the losses associated with traversing the hole, and present results for a range of different shapes of hole. We then discuss the implications of our results with respect to the prospects of reaching the strong coupling regime for atoms confined in such a hole. We also examine the intensity distribution of the light inside the hole and use our results to show that both crossed and individual waveguides are capable of forming a full, 3D dipole trap for ultracold atoms using only guided light.

While this article focuses on the specific application of a light-atom interface, the results may also be applicable within other fields. These include fibre-based gas sensors \cite{fibgasen1,fibgasen2} and the efficient transmission of light through integrated optical or optofluidic elements \cite{intel1,intel2} in a fibre or waveguide chip. Coupling of light between waveguides and/or optical fibres is also an important area of application. 

\section{Simulation methods}

In order to identify a geometry which allows a gap in the micrometer range with high optical transmission, simulations based on solving Maxwell's equations were performed using Optiwave software (Optiwave Systems Inc.). For most data we use a three dimensional finite difference time domain (FDTD) method, which is a numerical solution of the full Maxwell's equations \cite{optifdtd}.

Where applicable, we compare that to the results of a beam propagation method (BPM). While numerically less intensive than the FDTD method, the BPM we employ makes several approximations --- most notably the paraxial approximation --- that render it less accurate than the FDTD method. We therefore use it only as a guide to the overall trends in the system's behaviour, which helps to highlight the most interesting areas for investigation with the more time-consuming FDTD simulations.

The key parameters of the FDTD simulation method are the mesh spacing in each dimension $\Delta x,\Delta y, \Delta z$ (with the $z$ axis corresponding to the direction of light propagation along the waveguide), the time step size $\Delta t$ and the number of time steps for which the simulation was run, $N_t$. The parameters used for each simulation are given in supplementary table 1. The finite size of the spatial and temporal discretisation inevitably leads to a numerical error in the results of the simulations, an estimate of which is shown as error bars. Full details of how we estimate the magnitude of this error are given in the supplementary material. The boundary condition used for the FDTD simulations was the built-in anisotropic, perfectly-matched layers (PML) condition, which accurately approximates a perfect absorber.

In general we consider the overlap of the transmitted light with the fundamental mode of the waveguide (MOTL), rather than the transmission coefficient. The difference between these two is the additional loss of light resulting from reflections at the glass to air/vacuum interfaces. In the case of rectangular holes interference effects are important, and the reflection coefficient varies strongly with the length of the hole. We therefore calculated reflection coefficients for several illustrative cases of rectangular hole. The maximum and minimum (excluding sub-wavelength holes where the reflection coefficient tends to zero as the length tends to zero) were 17.8\% and 0.5\%, occurring for holes with lengths of 4 and 1.7 micrometers respectively. Reflection coefficients for rectangular holes with lengths of 1, 5, 12, 22 and 30 micrometers were found to be 15.9, 1.6, 14.8, 15.6 and 11.9\% respectively.

It is also worth noting that for rectangular holes the reflected light typically overlaps well ($\sim$97\% intensity overlap) with the guided mode of the waveguide. Therefore, if such a hole were placed inside an optical resonator, it would not be accurate to regard the majority of the reflected light as lost from the system.

However, calculation of accurate reflection/transmission coefficients requires a long simulation time, to allow for multiple reflections within the system.  When the surfaces of the hole have even modest curvature, the poor overlap and phase averaging between the light reflected via different pathways means that interference effects are essentially negligible, and the reflection losses at each interface can be treated as independent. This yields reflection losses of $\sim4$\% per interface, with little variation according to the parameters of the hole. In these cases the MOTL is therefore the most interesting system property to consider. 





With a specific application in mind, i.e. the interaction of photons with Caesium atoms, we consider the case of 852 nm light (resonant with the D$_2$ line in Caesium) in a waveguide whose refractive index profile matches a commercial optical fibre (Thorlabs 780HP), as a representative example for a typical singlemode optical waveguide. However, the general trends and qualitative behaviours observed are likely to be widely applicable. We also consider a specific case based on the waveguide chip described in \cite{hinds}, for which we find agreement between our simulations and the work of the original authors.

\section{Simulation results}
\label{results}

\begin{figure}
 \begin{center}
\includegraphics[width = 14cm]{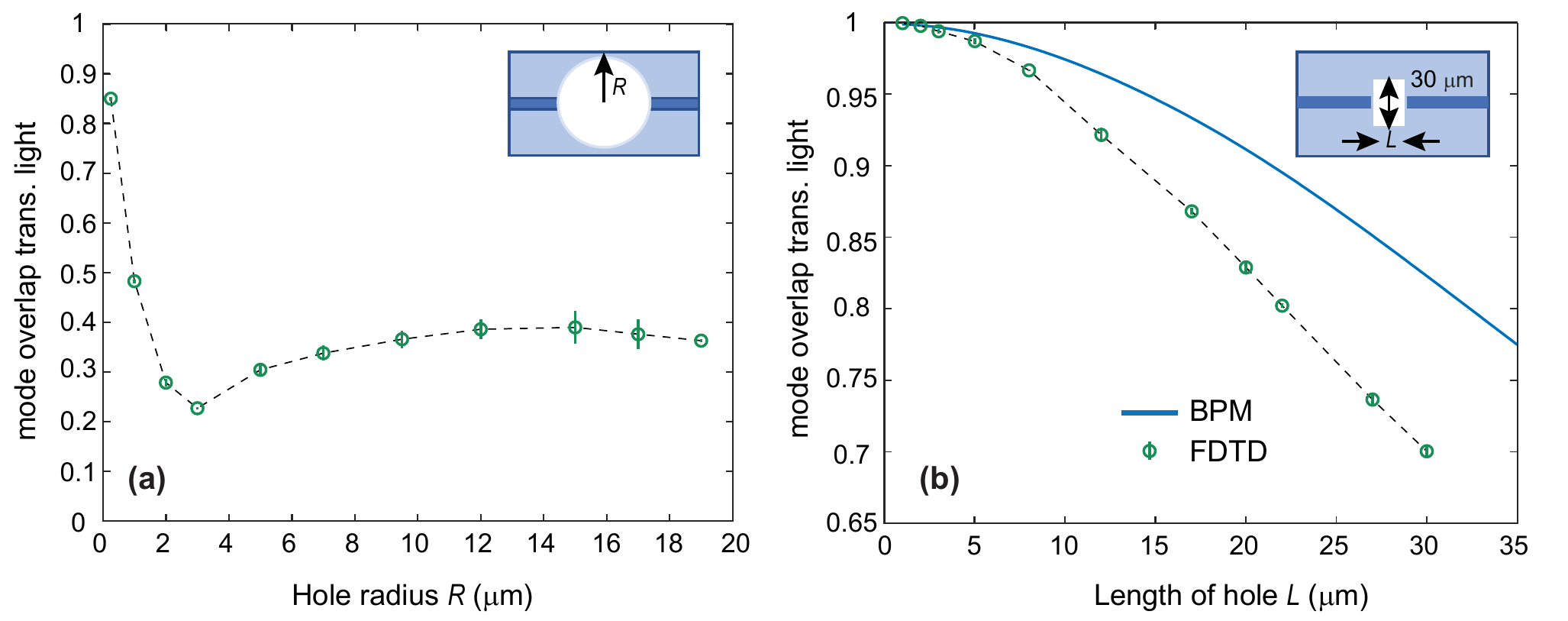}
 \caption{Fractional power overlap with the fundamental mode of the waveguide for light transmitted across a cylindrical (a) and a rectangular (b) through-hole in an optical fibre as a function of hole length. The hole is orientated perpendicular to the fibre axis, with the insets displaying the considered geometry. In both cases, the distance of closest approach between the input and output surfaces is 30 $\mu$m, the fibre parameters used match Thorlabs 780HP fibre, and the optical wavelength is 852 nm.}
 \label{cyl_hole}
 \end{center}
 \end{figure}

The first hole shape considered is the simplest geometry  --- a cylinder, and the results are plotted in figure \ref{cyl_hole}(a). The dip in MOTL between 10 and 1.5 $\mu$m is due to additional divergence caused by the concave surface curvature, and MOTL eventually tends upwards to 1 as the hole size is reduced to zero. For holes with diameters greater than about three micrometers, the highest achievable MOTL is $\sim$39\%. This occurs at a plateau of MOTL as a function of diameter, for hole diameters from 20 to 40 micrometers.

The next case considered was a rectangular hole, as the flat faces of the rectangular hole were expected to eliminate the dip at small radii seen in the cylindrical hole due to concave lensing effects. This was indeed found to be the case, and the results are plotted in figure \ref{cyl_hole}(b).

The FDTD and BPM results show a similar trend, where the FDTD result is consistently lower, owing to the remaining inaccuracy of the BPM data due to the paraxial approximation not covering all beam paths in this regime. For holes up to 10 optical wavelengths, a mode overlap $> 95$\% is achievable. 

For rectangular holes some experimental data is available in the literature \cite{hinds}. This was found to yield 65\% power transmission across a 16 $\mu$m gap between 4 $\mu$m square waveguides in silica. Based on the refractive index contrast (0.75\%) and laser wavelength used, we simulated this situation using the FDTD method and found an estimated transmission of $\sim 78$\% (including reflection losses). The theoretical result is an upper bound to what is achievable experimentally and the slightly lower result can be explained by potential imperfections in the polishing of the end facets, a small angle between the end facets or a small amount of contamination. We therefore find that these figures are consistent with our expectations.

From the two initial simulations, it was expected that the use of convex surfaces should enhance the mode overlap for larger holes, as the focal power of these surfaces will compensate for the beam divergence related to the numerical aperture of the fibre core and allow recapture of light into the guided mode.

\begin{figure}
 \begin{center}
\includegraphics[width = 14cm]{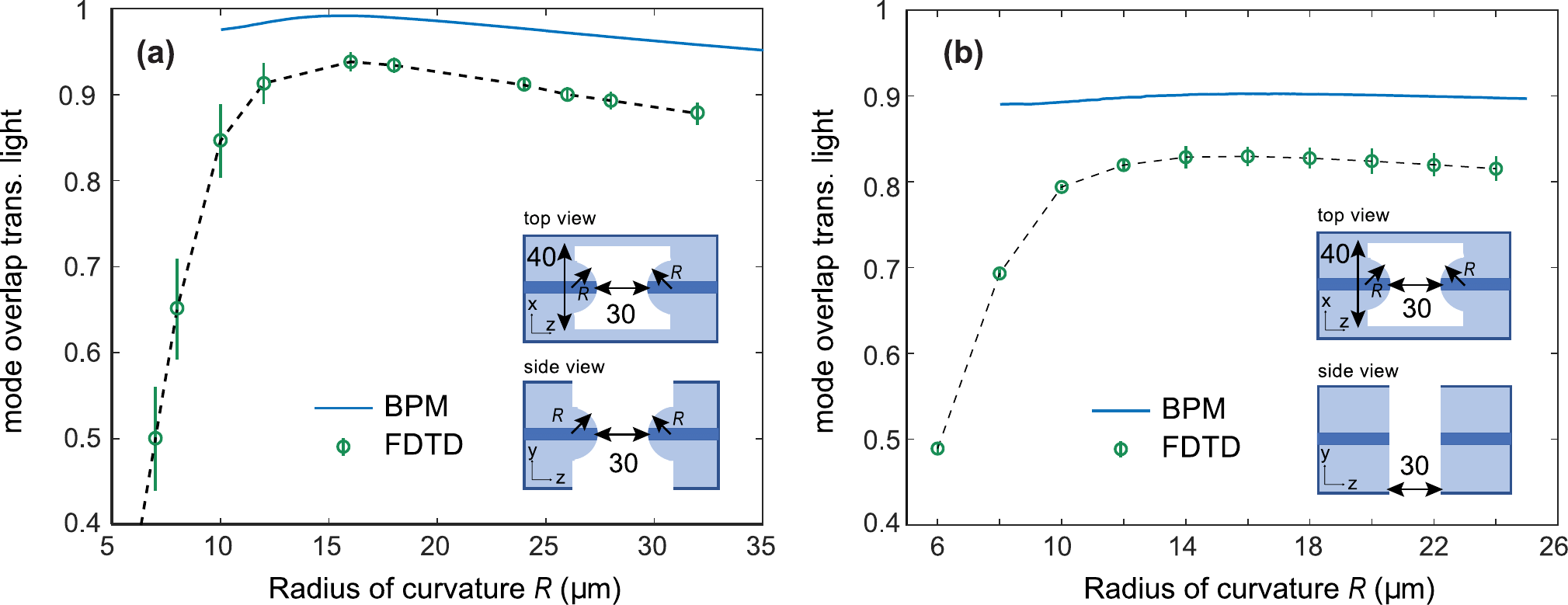}
\caption{(a) Fractional power overlap with the fundamental mode of the waveguide for light transmitted across a through-hole in an optical fibre with spherically-curved convex  input and output facets, as a function of the radius of curvature of the surface. (b) The same for cylindrically-curved input and output facets. In both cases, the distance of closest approach between the input and output surfaces is 30 $\mu$m, the fibre parameters used match Thorlabs 780HP fibre, and the optical wavelength is 852 nm.}
 \label{sphere_hole}
 \end{center}
\end{figure}

One relevant case to consider is that of a hole with spherically-curved surfaces on the input and output facets. We simulated this case for a range of radii of curvature, with the closest approach between the input and output surfaces being locked to 30 $\mu$m. The results are plotted in figure \ref{sphere_hole}(a). It can be seen from the FDTD simulations that for a radius of curvature of $\sim$16 $\mu$m the MOTL exceeds 93\%, a major improvement over the rectangular hole of 30 $\mu$m length where the MOTL was only $\sim70\%$ (see figure \ref{cyl_hole}(b)). As a consistency check, additional simulations were also run for radii of curvature up to 3 mm, well beyond the range of figure \ref{sphere_hole}(a). It was confirmed that, as expected, the MOTL gradually drops off as the radius of curvature is increased and ultimately tends to the same value predicted for the 30 $\mu$m rectangular hole.  


With the practicalities of making a hole of this shape in mind, we also considered the effect of using cylindrically curved convex surfaces instead of spherically curved surfaces. We expect such holes to be easier to make as they have a constant cross-section. The results are shown in figure \ref{sphere_hole}(b).  



\begin{figure}
 \begin{center}
 \includegraphics[width = \textwidth]{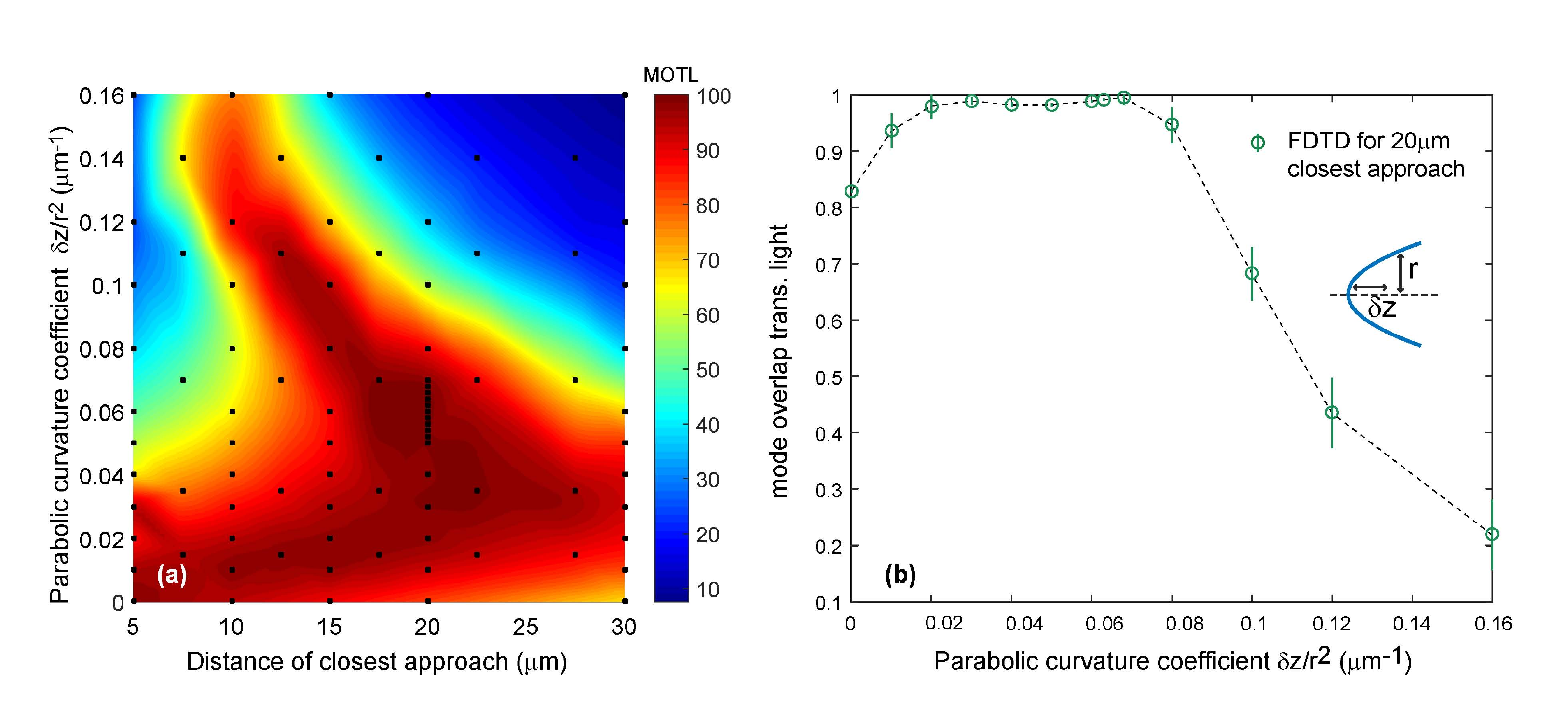}
 \caption{(a) Mode overlap for light transmitted across a through-hole in an optical fibre with parabolically-curved convex input and output facets, as a function of the coefficient of curvature of the surfaces and the distance of closest approach of the input and output faces. (b) MOTL as a function of the coefficient of curvature with the distance of closest approach equal to 20 $\mu$m.  In (a) the black markers indicate the locations of the data points resulting from FDTD simulations, while the colour map is a cubic interpolation between the measured points.}
 \label{parabolic}
 \end{center}
\end{figure}

Parabolic surface curvatures were also investigated using the FDTD method, and were found to provide even better mode overlap for the transmitted light. The results are plotted in figure \ref{parabolic}(a). In particular, for circularly symmetric, convex, parabolic surface curvature of the hole surfaces with $\alpha = \frac{\delta z}{r^2} = 0.068~\mu$m$^{-1}$ we find $(99.5^{+0.5}_{-1.3})$\% MOTL for a hole where the distance of closest approach is 20 $\mu$m.

The convex surfaces also lead to focusing of the light within the hole. The enhancement in peak intensity that results from this could be useful for the production of an optical dipole trap within the hole (see below), or to allow the strong coupling regime to be reached with smaller atomic ensembles as discussed below. We observe enhancements of up to a factor of 15.5 in peak intensity, with a general trend towards greater surface curvature producing a larger peak intensity enhancement.  In the example given above (parabolic curvature with $\alpha = 0.068~\mu$m$^{-1}$) we find that the peak intensity is increased by a factor of $\sim$5.

In order to constitute a quantum memory, or to allow longer interrogation times in sensing applications, it is advantageous to hold cold atoms within the junction, e.g. in an optical dipole trap. Where in other systems the creation of a small stable trap could be challenging, here they can be created using only light guided in the waveguides themselves. The trapping region then also automatically overlaps with the interrogation region of the probe light. 

\begin{figure}
 \begin{center}
 \includegraphics[width = \textwidth]{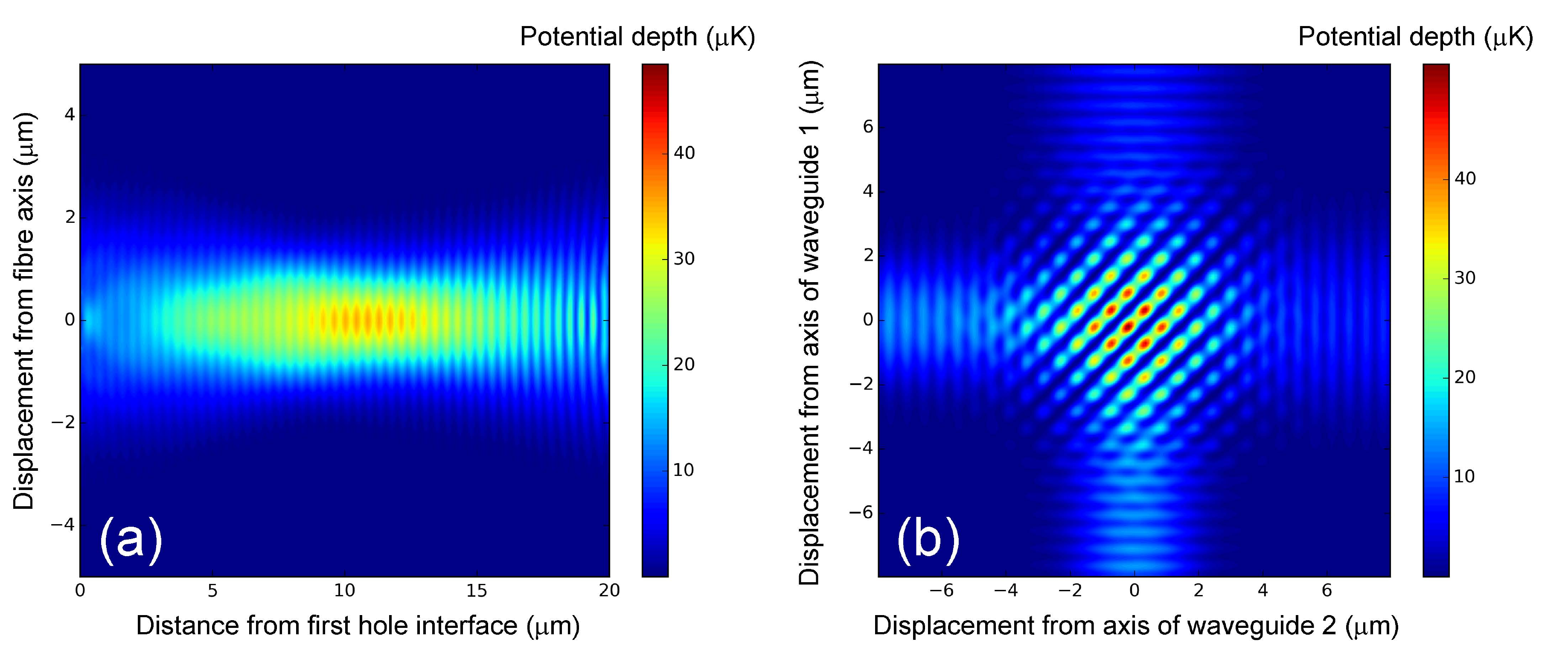}
 \caption{Simulated trapping potentials for Cs atoms in two example waveguides. Part (a) represents a 20 $\mu$m long hole with convex parabolic surface curvature ($\alpha = \frac{\delta z}{r^2} = 0.063~\mu$m$^{-1}$) in a waveguide with parameters matching Thorlabs 780 HP optical fibre. Part (B) represents the intersection of two 4 $\mu$m square silica waveguides with a refractive index contrast of 0.75\%. In both cases each waveguide carries 1 mW of 1064 nm light propagating in the positive $x$ or $z$ direction as appropriate.}
 \label{dipole_pot}
 \end{center}
\end{figure}

The FDTD method of simulation provides full data on the electric field as a function of position within the junction and can therefore be used to determine the light intensity, and hence the optical dipole potential, as a function of position within the junction. Figure \ref{dipole_pot}(a) shows the dipole potential generated for Cs atoms by 1 mW of light at 1064\,nm crossing a 20 $\mu$m hole with convex parabolic surface curvature ($\alpha = 0.063~\mu$m$^{-1}$) in a waveguide with parameters matching Thorlabs 780 HP optical fibre.  It can be seen from this that already a single beam forms a full 3D trap - a result of focusing of the guided light at the convex interfaces. Figure \ref{dipole_pot}(b) shows the dipole potential generated in a junction formed at the intersection of two 4 $\mu$m square silica waveguides with a refractive index contrast of 0.75\%. There is assumed to be 1 mW of 1064 nm light in each waveguide, with identical linear polarisations. It is worth noting that the small mode area offers an unusually high trap depth for a given optical power and wavelength.   

The damage threshold for waveguides and optical fibres of this type is typically of the order of 10$^{10}$ Wm$^{-2}$ for visible and NIR wavelengths. This would correspond to a power of $\sim$200 mW in each waveguide. As a result, the maximum trap depth that can realistically be achieved in such a junction (with a comparable detuning of the trapping laser from the relevant atomic transition) would be on the order of 10 to 15 mK. 

\section{Strong coupling regime}
\label{scoup_theory}

Reaching the strong coupling regime is of interest as it permits single-photon gate operations and allows the observation and exploitation of quantum electrodynamic effects \cite{expcav1,expcav2,expcav3,theocav1,theocav2}. In order to reach this regime an optical cavity would be produced around a quantum system with an appropriate optically-addressable transition, in this case via the use of laser-written Bragg gratings \cite{expcav1,laser_bragg} on either side of the hole. Different regimes can then be reached, depending on the cavity length and the choice of hole type. 
The strong coupling regime is defined as the regime in which the atom-cavity coupling constant $g_1$ significantly exceeds the atomic decay rate $\gamma$ and the cavity decay rate $\kappa$. The Purcell regime, in which the cooperativity $C = g_{1}^2/(\kappa \gamma)$ is large but $g_{1}<\kappa$, is also of interest --- particularly with regard to the production of single photon sources \cite{sps}. For completeness, note that there are a few alternative definitions of $C$ in the literature. The maximum value of the atom-cavity coupling constant is given according to \cite{Reiserer2015}
\begin{equation}
 g_{1} = \zeta \sqrt{\frac{\omega_c}{2 \hbar \epsilon_0 V}} \phi (\vec{r}),
\label{coupstrength}
\end{equation}
where $\zeta$ is the dipole matrix element for the atomic transition being adressed, $\omega_c$ is the resonant frequency of the cavity and $V$ the volume of the cavity mode \cite{strng_cpl}. The mode volume is defined such that $V = \int \phi^2 (\vec{r}) d^3\vec{r}$. 

Due to the numerical intensity of the simulations required, we do not perform FDTD simulations including an optical cavity. Instead, loss of light from the guided mode on traversing a microscopic hole is modelled as the introduction of an additional intra-cavity loss mechanism. In this case the decay rate $\kappa$  is given by \cite{Reiserer2015,nanofib2,saleh}:
\begin{equation}
\kappa = \frac{c(1-T\sqrt{R_1 R_2})}{2 l_{c} \sqrt{T}(R_1 R_2)^{1/4}},
\label{lossrate}
\end{equation}
where $l_{c}$ is the optical path length of the cavity,  $R_1$ and $R_2$ are the mirror reflectivities and $T$ is the transmission coefficient past the intra-cavity loss source as determined by the FDTD simulations described above. 

We assume a $1/e^2$ intensity radius of $2.5~\mu$m, which is representative of most waveguides of the kind we consider herein, and analyse examples for different cavity lengths. In \cite{expcav1} fibre Bragg gratings with a reflectivity of $99.5\%$ were used, and even higher reflectivities are achievable \cite{expcav1, expcav2}.
For now we assume that the mode profile in the hole does not differ significantly from that in the waveguide, although the focal effects of surfaces with convex curvature discussed in the previous section could in principle be used to enhance the coupling strength.


Consider the case of a single Caesium atom addressed on the $D_2$ line $(F=4, m_F=4 \rightarrow F'=5, m_F'=5)$, trapped inside a rectangular void with a length of $L= 5~\mu$m inside an optical cavity with length $l_c=5~$mm. This situation was found to yield a MOTL of (98.7$\pm$0.3)\%, and applying equations (\ref{coupstrength}) and (\ref{lossrate}) therefore gives values of $g_{1}=187~$MHz and $\kappa=(552 \pm 100)~$MHz respectively. Here $g_{1}$ clearly exceeds the atomic decay rate of $\gamma=16.4~$MHz \cite{steck} (where $\gamma$ is equal to half of the spontaneous decay rate $\Gamma$, to allow for the atomic population distribution) and the cooperativity equates to $C=3.9 \pm 0.8$, thus placing this system within the Purcell regime \cite{purcell}. Note that for rectangular holes we consider only losses associated with imperfect MOTL. This is because, in the case of rectangular holes, the reflected light was found to overlap well with the mode of the waveguide \emph{and} the reflection coefficient could be reduced to $\sim$0.5\% through appropriate local tuning of the hole length. When considering holes with other shapes reflection losses are taken into account.

In order to enter the strong coupling regime it is also necessary that $g_{1} > \kappa$. This would be reached for a single Cs atom in a rectangular void of $L=5~\mu$m and a cavity length of $l_c=50~$mm, again for a cooperativity of $C=3.9 \pm 0.8$ and a coupling rate of $g_{1}=59~$MHz and $\kappa = (55 \pm 10)~$MHz. For longer rectangular holes, up to a size of $L=8~ \mu  m$ ($(96.7\pm 0.3)\%$ MOTL), cooperativities with $C=1.8 \pm 0.2$ and a cavity length of $l_c=300~$mm with $g_{1}=24~$MHz and $\kappa=(20.0 \pm 1.5)~$MHz are possible. 
While introducing a large ensemble of cold atoms into such a space is difficult, introducing a single trapped atom into a hole of this size is plausible, and small rectangular holes may therefore permit strong coupling of single atoms to guided light. 

Holes with convex surface curvatures may allow the strong coupling regime to be reached for even larger holes, due to a combination of increased MOTL and local field enhancement by the focal effects of the surfaces. For example, we find that convex, parabolic surface curvature with a coefficient of  $\frac{\delta z}{r^2} = 0.068~\mu$m$^{-1}$ allows a cooperativity of 4 to be achieved for a single Cs atom in a 20\,$\mu$m long hole. See supplementary material for full details.

When multiple atoms are trapped inside a void and therefore confined in such a cavity, there is a collective enhancement of the coupling constant by a factor equal to the square root of the number of atoms present $g_{N}=\sqrt{N}g_{1}$ \cite{collect_enhance, collect_enhance_rberg}, assuming the atoms all couple equally to the optical field. Considering the trapping volumes involved ($\sim 1 - 1000~\mu m^{3}$) and the densities typically achievable in a dipole trap ($\sim 1~\mu m^{-3}$ without evaporative cooling or $\sim 1000~\mu m^{-3}$ with \cite{dip_dense}), atom numbers from 1 to 10$^6$ should be achievable, with a corresponding increase in the achievable cooperativity. 
Calculation of an exact value for the collective cooperativity $C_N$ requires specification of both the number and distribution of atoms within the hole. As an example, confinement of 260 Cs atoms within a hole of length 20 $\mu$m with convex, parabolic surface curvature with a coefficient of  $\frac{\delta z}{r^2} = 0.068~\mu$m$^{-1}$ could be expected to yield cooperativities on the order of $C_N=400$, subject to reasonable assumptions about the distribution of the atoms within the hole. See supplementary material for full details.

Other quantum systems such as quantum dots \cite{computing_with_qdots} or semiconductor vacancy centres \cite{NVs} could be placed into much smaller holes, and potentially even into holes which are then filled with index-matching fluid, as is done in \cite{index_match}. Bringing these systems into the strong coupling or Purcell regimes should therefore also be possible in microscopic voids of this kind. 


\section{Outlook}
\label{outlook}

The transmission of light across holes in optical waveguides, with sizes in the range 1 to 30 micrometers, has been studied via numerical simulation. Most attention was given to losses resulting from mode mismatch, since for curved surfaces reflection losses remain roughly constant, at about 8\% for two uncoated glass surfaces. The results are found to be consistent with previous experimental results and reproduce the correct limiting behaviour as variables become large or small and calculation of the transmission becomes trivial.  

The results show that for a given length of hole the losses due to mode mismatch can be greatly reduced through appropriate hole shaping, with appropriate convex parabolic curvature of the input and output faces of a 20 micrometer long hole reducing losses due to mode-mismatch from $\sim$15\% in the case of flat end faces to $0.5(+1.3/-0.5)$\%. Shaping of holes to maximise power transmission may have applications in fibre-based sensors as well as in quantum optics experiments involving cold atoms. 

It is also shown that dipole traps for ultracold atoms can be formed within such holes using only guided light, with maximum depths in the range of ten to fifteen mK for typical silica waveguides and trapping laser detunings of $\sim$200 nm. Furthermore, our calculations suggest that construction of optical cavities around such holes should allow the strong coupling regime to be reached for single atoms trapped in holes up to $\sim$20 $\mu$m in length. Exploiting collective enhancement to increase the coupling strength allows the use of larger voids and permits higher cooperativities. This could make holes of this kind an ideal component for interfacing light and cold atoms as part of an integrated quantum information system.



Note that the potential effects of imperfect fabrication are not accounted for in our simulations. Future work will include determining suitable methods for the smoothing and anti-reflection coating of the interior hole surfaces. For example, smoothing is expected to be possible using either ion beam milling or plasma assisted chemical etching \cite{ionmill}. Additionally, shaping of the waveguide's refractive index profile on either side of the junction will be investigated. Previous experimental work has demonstrated the plausibility of shaping such waveguides \cite{shapeguides1,shapeguides2}, and the additional dimension this approach adds to the space of free parameters available when designing a waveguide-void interface should allow for extremely high transmission coefficients to be achieved.



\section{Data Availability Statement}

Any relevant data not presented in the manuscript is available from the authors upon reasonable request.

\section{References}

\bibliographystyle{unsrt}
\bibliography{fbrbib}



\section{Acknowledgments}

This work was supported by the EPSRC grants EP/R024111/1 and EP/M013294/1 and by European Comission grants QuILMI (no. 295293) and ErBeStA (no. 800942). The authors would like to thank Joerg Goette for useful discussions.

\section{Author contributions}

L.H. and N.C. provided the initial ideas for the work. C.B. commenced the simulation work, which was then significantly expanded by N.C. and E.D.. M.T.G. provided advice on the efficient use of the simulation software. E.D., V.N. and L.H. assisted N.C. with manuscript preparation and review of the relevant literature. All authors reviewed the final manuscript.

\section{Additional information}

The authors declare no competing interests.

\end{document}


\maketitle

All numerical simulation methods have an associated numerical error. The magnitude of this error can be estimated via convergence testing of the simulation results. This supplementary information gives details of the convergence testing performed on the simulations used in the main article and explains how the results of this testing were used to estimate the numerical errors on the data shown. It also contains a table giving the simulation parameters used in all of the simulations described in the main article.

\section{FDTD convergence testing}


Direct testing confirmed that the level of discretisation of the spatial mesh was the dominant source of inaccuracy in our simulations, with variations in output for a given fractional change in mesh spacing greatly exceeding those for the same fractional changes in the time step increment or the dimensions of the volume included in the simulation. 

The memory and computation time requirements of an FDTD simulation are strongly dependent on the number of spatial mesh cells used. If the number of mesh cells used along each axis is scaled by a factor $\xi$, then the memory requirement scales as $\xi^3$ and the simulation runtime as $\xi^4$ \cite{requirements}. This scaling placed limits on how fine a spatial mesh could reasonably be used in these simulations, and an estimation of the remaining numerical error was performed. The first step towards obtaining these estimates was to study the convergence behaviour of the simulations in detail for some representative example cases. 

For each fibre-hole geometry at least one test of the convergence behaviour of the simulations was carried out, based on the response of the simulated MOTL to variations in the size of the spatial mesh used for the simulation. To do this, we define a mesh scaling factor $F$ such that, if our original simulation was run with the spatial mesh $(\Delta x,\Delta y,\Delta z) = (a,b,c)$ then a simulation with mesh scaling factor $F$ would use $(\Delta x,\Delta y,\Delta z) = F(a,b,c)$. We then run multiple simulations with different values of $F$ and study the corresponding variation of the apparent MOTL. The results are shown in supplementary figure \ref{all}.

\begin{figure}
 \begin{center}
 \includegraphics[width = \textwidth]{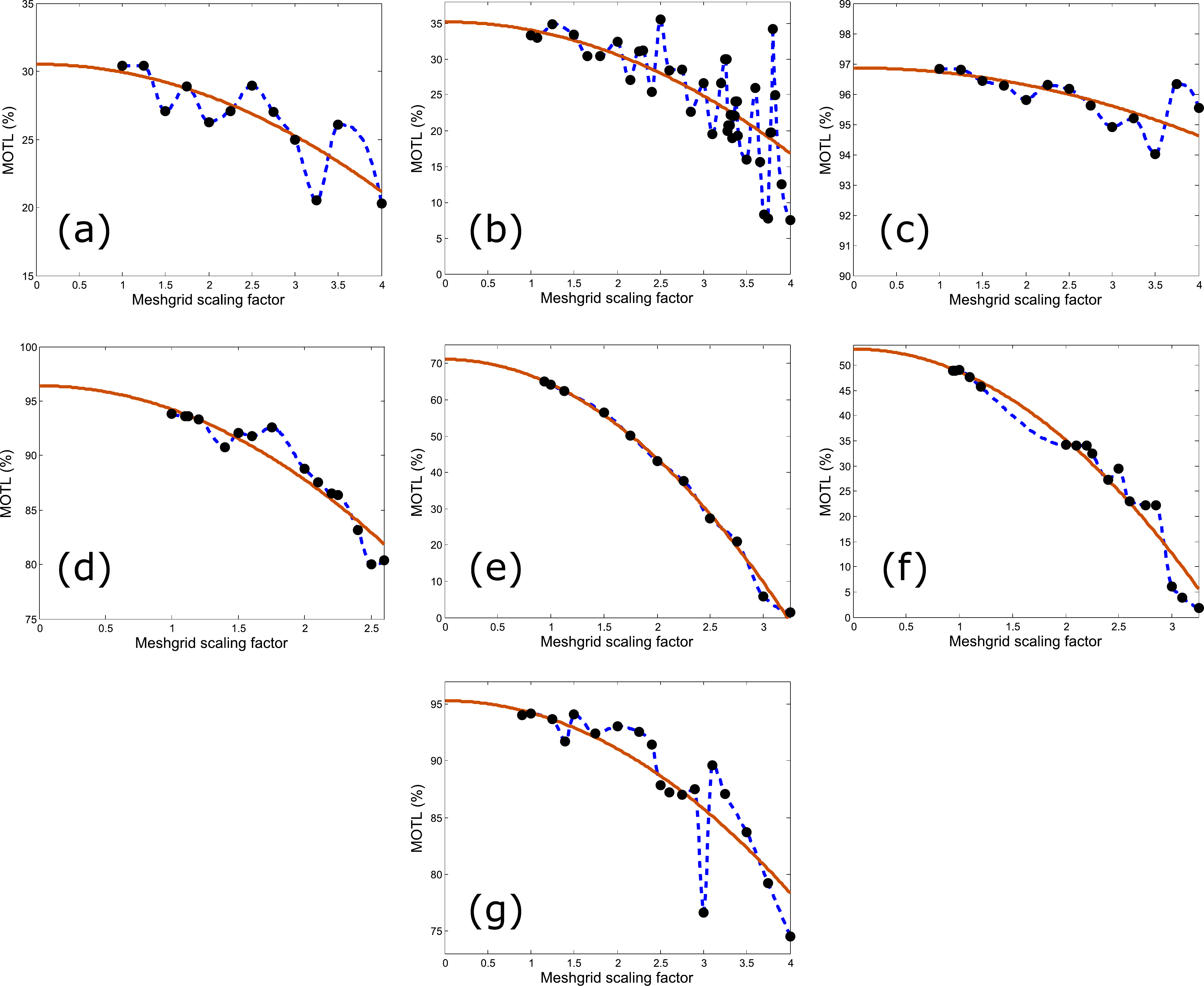}
 \caption{Convergence testing results. The plots show simulated MOTL as a function of the mesh spacing scaling factor $F$, such that the mesh spacing for any given point is equal to the base mesh spacing multiplied by $F$. The orange line is a quadratic fit to the data, assuming errors proportional to $F^2$ on the individual data points.  Panels (a) and (b) show results for holes with cylindrical, concave curvature (corresponding to figure 2(a)) with radii of 5 $\mu$m (a) and 7 $\mu$m (b). The base mesh spacing was $(\Delta x, \Delta y, \Delta z) = (0.04,0.08,0.04)~\mu$m for panel (a) and $(\Delta x, \Delta y, \Delta z) = (0.07,0.07,0.047)~\mu$m for panel (b). Panel (c) shows results for a rectangular hole (figure 2(b)) of length 8 $\mu$m, with a base meshgrid of $(\Delta x, \Delta y, \Delta z) = (0.08,0.08,0.04)~\mu$m. Panels (d) and (e) show results for convex, spherical surface curvature (figure 3(a)). The radius of curvature is 8 $\mu$m in (d) and 16 $\mu$m in (e). The base mesh spacing was $(\Delta x, \Delta y, \Delta z) = (0.08,0.08,0.058)~\mu$m in both panels. Panels (f) and (g) show results for holes with convex, cylindrical curvature (figure 3(b)) with a radius of 6 $\mu$m (f) and parabolic curvature (figure 4(b)) with a curvature coefficient of 0.01 $\mu$m$^{-1}$ (g). The base mesh spacing was $(\Delta x, \Delta y, \Delta z) = (0.08,0.08,0.058)~\mu$m for panel (f) and $(\Delta x, \Delta y, \Delta z) = (0.08,0.08,0.08)~\mu$m for panel (g). }
 \label{all}
 \end{center}
\end{figure}

Numerical inaccuracies resulting from spatial mesh size are known to have a second order (and higher) dependence on the mesh size, resulting from the use of the central difference approximation \cite{optifdtd_basics,error_scaling}. The exception to this is ``staircasing'' errors for curved surfaces, which can cause inaccuracies with a first order dependence on mesh spacing \cite{staircasing}. However, in the scenarios we consider, the characteristic length scale of the physical structures involved is typically much larger than the optical wavelength, suggesting that staircasing is unlikely to be the dominant source of error. It can be seen from supplementary figure \ref{all} that the apparent MOTL is well approximated as a quadratic function of the mesh scaling factor, in some cases with additional, oscillatory behaviour that is itself bounded by an envelope with a quadratic dependence on the mesh scaling factor. This is consistent with our expectations. We do not find evidence of a linear dependence of the apparent MOTL on mesh spacing, giving a further indication that staircasing effects do not contribute significantly to our numerical error.

\section{Error estimation}

Based on the results above, the assumption of quadratic scaling of the numerical error with the mesh size was used to determine the magnitude of our numerical errors. As error estimates are required for all data points, a faster method than full convergence testing was used. By determining the simulated MOTL for two different mesh spacings and extrapolating back to zero mesh spacing (assuming variation in apparent MOTL proportional to $F^2$), it was possible to estimate the difference between the quoted results and an extrapolated `perfect' simulation with zero mesh spacing. This allows bounds to be placed on the deviation of our results from the true values (see supplementary figure \ref{twopoint_method}). The resulting error estimate for an individual data point is then given by:
\begin{equation}
 \mathrm{Error} = \frac{\left|\mathrm{MOTL}_1-\mathrm{MOTL}_2\right|}{F_2^2-F_1^2}F_1^2,
\end{equation}
where $F_1$ and $F_2$ are the scaling factors for the size of the two different spatial meshes used and MOTL$_1$ and MOTL$_2$ are the corresponding simulated MOTL values. The presence of higher order terms in the function describing the relationship between the mesh spacing and the numerical inaccuracy will on average lead to an overestimation of the error when applying equation (1), not an underestimate. 

\begin{figure}
 \begin{center}
 \includegraphics[width = \textwidth]{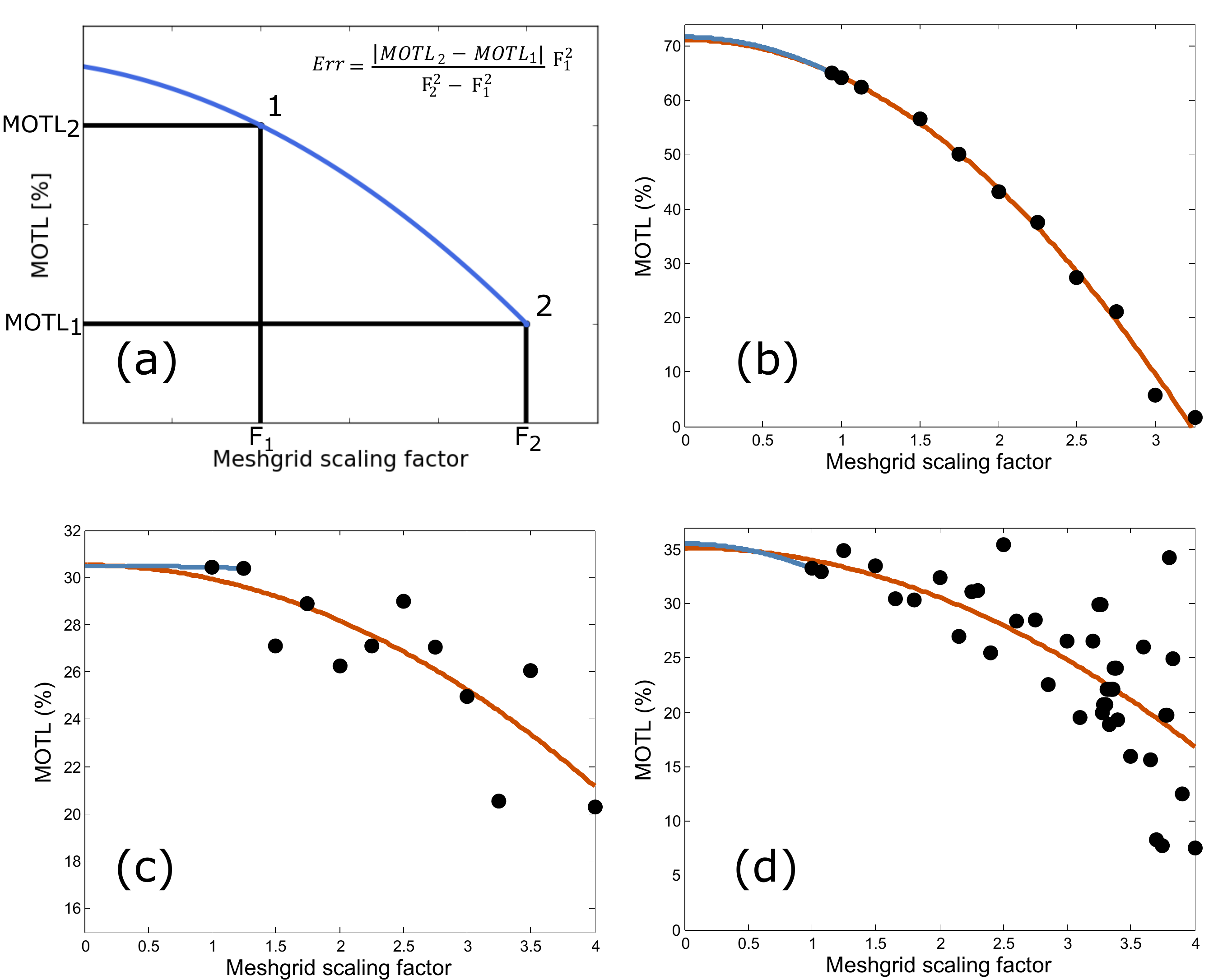}
 \caption{(a) Sketch indicating the method of determining an error estimate from two simulations with different meshgrids. (b)---(d) Examples of implementation of this method for the cases presented in supplementary figures \ref{all}(e), \ref{all}(a) and \ref{all}(b) respectively. The blue line represents a fit to the two points with the smallest meshgrid, as used to determine the error, while the orange line represents a fit to all of the data resulting from the full convergence test. Panel (b) shows excellent agreement between the two methods, while panel (c) shows the two-point method leading to an underestimate of the error and panel (d) shows an overestimate. Since we always consider only the magnitude of the error resulting from these estimates (and neglect the sign), it can be seen that the deviations of the two-point method from the many-point fit, caused by local fluctuations in the simulated MOTL, will on average lead to an overestimation of the error rather than an underestimation. To reduce the influence of these fluctuations, we also apply a moving average to the error estimates made for points taken with the same meshgrid and sample geometry.}
 \label{twopoint_method}
 \end{center}
\end{figure}

Figure \ref{twopoint_method} shows examples of the application of this method, for which we see good agreement between this method and one based on fitting all of the available convergence data. However, the applied method is more susceptible to `noise' arising from short-scale fluctuations in apparent MOTL as a function of $F$, which can affect the quality of the error estimate. In order to mitigate this effect, a 3 point moving average was applied to the numerical error estimates for like points (same meshgrid and geometric form). The resulting estimates for the bounds on our numerical uncertainty are plotted as error bars in the figures in the main article. 

Local fluctuations in apparent MOTL occur on a length scale that is short compared to the base mesh spacing of the simulations, with no discernible long-distance correlations. Since we always use the absolute value of (MOTL$_1 - $MOTL$_2$) in our error estimates, they will on average lead to an overestimate of the numerical error rather than an underestimate. As previously discussed, the same is true for the presence of higher order terms in the scaling of the numerical inaccuracy with mesh size. Consequently the error obtained from equation (1), together with local averaging, leads to a slight overestimation of the simulation uncertainty. 


\section{Atom-light coupling strength in holes with convex surface curvature}

Convex surface curvatures can allow high cooperativities even for larger holes. There are two reasons for this: the increased MOTL that can be achieved and a focusing effect that leads to a local enhancement of the strength of the optical field in the centre of the hole, corresponding to an increase in the value of $\phi (r)$ in equation (1) of the main article. 

One example considered is a hole of length 20 $\mu$m with convex, parabolic surface curvature with a coefficient of  $\frac{\delta z}{r^2} = 0.068~\mu$m$^{-1}$. In this case the maximum local field strength is increased by a factor of 2.3 by the focal effects of the surfaces. The MOTL value determined for this scenario by our simulations is $(99.5^{+0.5}_{-1.3})$\%. Allowing for 8\% reflection losses the corresponding values of $g_{1}$ and $\kappa$ are 55 MHz and (47$^{+7}_{-3}$) MHz respectively. This yields $C=4.0^{+0.2}_{-0.5}$. 

Furthermore, the development of an anti-reflection coating technique for the interior hole surfaces would greatly expand the range of options available. If reflection losses are neglected, the scenario above leads to predicted values of $\kappa=5.0$ MHz (including numerical error in MOTL up to 11.5 MHz) and $g_{1}=55~$MHz. The corresponding predicted cooperativity is C$=37$ (minimum of 16 with numerical MOTL error). 

Another example considered in the main article is the confinement of 260 Cs atoms within a hole of length 20 $\mu$m with convex, parabolic surface curvature with a coefficient of  $\frac{\delta z}{r^2} = 0.068~\mu$m$^{-1}$. This could yield collective cooperativities anywhere between $C_N=98$ for a uniform atom distribution and $C_N=1040$ if all of the atoms could be localised at the exact point of maximum field strength. To reach the estimate of $C_N \sim 400$ given in the main text, we assume that the atoms could be localised to a region such that the local intensity enhancement of the optical field, averaged over the atoms present, was approximately a factor of 4. This would correspond to the case of atoms confined using a dipole trap, based on light guided through the waveguide itself as discussed in the main text, whose depth was roughly 5 times the characteristic thermal energy of the trapped atomic ensemble. This is a very reasonable figure given the trap depths considered in the main text and the typical temperature of Cs atoms after magneto-optical trapping (generally of the order of 100 $\mu$K). 

\section{Table of simulation parameters}

Supplementary table 1 shows the details of the simulation parameters used to produce all of the data displayed in the main text, including the secondary meshgrids used to estimate the numerical error.

\begin{table}
  \begin{center}
    \caption{Synoptic table stating all simulation parameters for the simulations presented in the main text of the paper. Figure labels including ``(secondary)" refer to the additional simulations used to estimate the numerical uncertainties.  The variable $R$ represents a radius of curvature or hole radius, $l$ the length of a rectangular hole and $k$ the curvature coefficient for parabolic surface curvature. All distances are given in micrometers and all times in attoseconds.}
    \label{tab:table1}
    \begin{tabular}{l|c|r} 
      \textbf{Main text} & \textbf{Subregion} & \textbf{Corresponding parameters}\\
      \textbf{figure}       &                                &  $(\Delta x,\Delta y, \Delta z, \Delta t, N_t)$\\      
      \hline
      2(a) & $R = 0.25~\mu$m &   (0.04, 0.08, 0.04, 66.7, 1680)	\\
      2(a) & $1~\mu$m $\leq R \leq 5~\mu$m &  (0.04, 0.08, 0.04, 66.7, 1890)\\
      2(a) & $R = 7~\mu$m &  (0.04, 0.08, 0.04, 67.1, 2058)		\\
      2(a) & $R > 7~\mu$m &  (0.08, 0.08, 0.0537, 99.6, 2542)	\\
      2(a) (secondary) & $R = 0.25~\mu$m &  (0.05, 0.1, 0.05, 83.4,1360)\\
      2(a) (secondary) & $1~\mu$m $\leq R \leq5~\mu$m &  (0.05, 0.1, 0.05, 82.4, 1496)\\
      2(a) (secondary) & $R = 7~\mu$m & (0.05, 0.1, 0.05, 83.4, 1666)	\\
      2(a) (secondary) & $R > 7~\mu$m & (0.085, 0.085, 0.0571, 95.3, 2378)	\\
      2(b) & $l \leq 5~\mu$m & (0.08, 0.08, 0.04, 66.7, 1764)	\\
      2(b) & $l > 5~\mu$m & (0.08, 0.08, 0.04, 66.8, 2898)	\\
      2(b) (secondary) & $l \leq 5~\mu$m & (0.09, 0.09, 0.045,75.1, 1591)	\\
      2(b) (secondary) & $l > 5~\mu$m &  (0.09, 0.09, 0.045, 75.1, 2590)	\\
      3(a) & $R < 10~\mu$m &  (0.075, 0.075, 0.0547, 91.3, 2542)	\\
      3(a) & $R = 10~\mu$m &  (0.08, 0.08, 0.0584, 50, 4592)	\\
      3(a) & $R > 10~\mu$m &  (0.08, 0.08, 0.0584, 97.4, 2349)	\\
      3(a) (secondary) & $R < 10~\mu$m & (0.08, 0.08, 0.0584, 97.4, 2407)	\\
      3(a) (secondary) & $R \geq 10~\mu$m &  (0.09, 0.09, 0.0657, 110, 2100)	\\
      3(b) & $R \leq 10~\mu$m &  (0.075, 0.075, 0.0547, 91.3, 2511)	\\
      3(b) & $R > 10~\mu$m &   (0.08, 0.08, 0.0584, 97.4, 2349)	\\
      3(b) (secondary) & $R \leq 10~\mu$m &  (0.08, 0.08, 0.0584, 97.4, 2349)	\\
      3(b) (secondary) & $R > 10~\mu$m &  (0.09, 0.09, 0.0657, 97.4, 2349)	\\
      4(a) & All &  (0.0836, 0.0836, 0.0836, 140, 2280)	\\
      4(b) & $k = 0$ &  (0.08, 0.08, 0.04, 66.8, 2898)	\\
      4(b) & $0 < k \leq 0.1~\mu$m$^{-1}$ &  (0.08, 0.08, 0.055, 91.8, 2790)	\\
      4(b) & $k > 0.1~\mu$m$^{-1}$ &  (0.085, 0.085, 0.0584, 97.5, 2987)	\\
      4(b) (secondary) & $k = 0$ & (0.09, 0.09, 0.045, 75.1, 2590)	\\
      4(b) (secondary) & $0 < k \leq 0.12~\mu$m$^{-1}$ &  (0.09, 0.09, 0.0619, 103, 2457)	\\
      4(b) (secondary) & $k > 0.12~\mu$m$^{-1}$ &  (0.09, 0.09, 0.0619, 103, 2808)	\\
    \end{tabular}
  \end{center}
\end{table}

\bibliographystyle{unsrt}
\bibliography{fbrbib}